\documentclass[letterpaper,11pt]{JHEP3}

\usepackage{axodraw,amsmath,epsfig}

\def\be{\begin{equation}}
\def\ee{\end{equation}}

\newcommand{\psl}{\!\not\! p}
\newcommand{\parsl}{\!\not\! \partial}
\newcommand{\chiR}{\chi_{_{\textstyle _R}}}
\newcommand{\chiRbar}{\bar\chi_{_{\textstyle _R}}}
\newcommand{\chiL}{\chi_{_{\textstyle _L}}}
\newcommand{\chiLbar}{\bar\chi_{_{\textstyle _L}}}

\preprint{UAB-FT-566}

\title{Holography for fermions}

\author{Roberto Contino \\ Department of Physics and Astronomy, Johns Hopkins University
Baltimore, Maryland 21218, USA \\ E-mail: \email{contino@pha.jhu.edu}}
\author{Alex Pomarol \\ IFAE, Universitat Aut{\`o}noma de Barcelona,
08193 Bellaterra (Barcelona), Spain \\ E-mail: \email{pomarol@ifae.es}}

\abstract{
The holographic interpretation is a useful tool to 
describe 5D field theories in a 4D language.
In particular it allows one to relate 5D AdS theories with 4D CFTs.
We elaborate on the 5D/4D dictionary for the case of fermions 
in AdS$_5$ with boundaries. This dictionary is quite useful
to address phenomenological issues in a very simple manner,
as we show by giving some examples.
}

\begin{document}

\section{Introduction}

Five-dimensional field theories with boundaries have received a lot of attention 
in the last years due to their possible applications to particles physics.
The historical approach to these theories has consisted in
decomposing the extra-dimensional fields in infinite series of 4D mass-eigenstates, 
the Kaluza-Klein (KK) modes, and calculate with them physical quantities.
Nevertheless, this KK approach is not always  very transparent, 
lacking sometimes of an intuitive  understanding of the  results.
This is especially true in the case of a warped extra dimension~\cite{Randall:1999ee}.

For this purpose the holographic or boundary procedure is much more useful.
It  consists in separating the bulk fields from their boundary value
and treating them as distinct variables. 
By integrating out the bulk, one obtains a 4D
theory defined on the boundary.
Two important benefits emerge from this approach.
First, since in most of the cases 
the bulk is weakly coupled to the boundary,   
one can treat the bulk as a small 
perturbation to the boundary action.
This can greatly simplify the calculations. 
An example is the one-loop running of the gauge coupling,
where the obscure KK calculation becomes 
strikingly simple if one uses the holographic approach~\cite{arpora}.
Second, one can observe that the effects of the bulk 
on the boundary fields resemble those 
 of a 4D strongly coupled sector.
This observation allows one to describe 5D theories
using a 4D language, from which we have a better intuition.
This is the basis of the AdS/CFT correspondence that has been conjectured 
in string theory~\cite{Maldacena:1997re}.
At the field theoretical level, however, we can consider holography 
as a tool to improve our understanding of 5D theories.

The holographic approach can be used to write a dictionary
relating 5D theories with 4D ones. In this article we study this 
correspondence in the case of 5D theories with fermions.
In the next section we present the holographic approach 
as applied to a general class of 5D theories.
This is based on the effective boundary action
that we compute for fermions in section~\ref{holoforfermions}. 
In section~\ref{AdSholo} we will
give the holographic interpretation for the case of fermions in
a slice of AdS$_5$ space. Some possible applications of holography
to phenomenology are given in section~\ref{pheap}.
We conclude by giving the holographic dictionary for fermions.

\section{The boundary action and holography}

Consider a five-dimensional theory with metric
\begin{equation}
ds^2=a(z)^2 \left( \eta_{\mu\nu} dx^\mu dx^\nu -dz^2 \right) \equiv g_{MN}\, dx^M dx^N \, ,
\label{metric}
\end{equation}
where  the fifth dimension $z$ is compactified on a  manifold with boundaries 
at $z=L_0$ and $z=L_1$ ($L_0\leq z\leq L_1$).~\footnote{Our notation is the following:
5D (4D) spacetime coordinates are labeled by capital Latin (small Greek) letters, 
$M=(\mu,5)$ where $\mu=0,\dots,3$ and $x^5=z$. We will also use capital (small) Latin letters 
to denote 5D (4D) Lorentz indexes, $A=(a,5)$.}
These boundaries will be called  ultraviolet (UV) boundary and 
infrared (IR) boundary respectively. We are interested in obtaining 
the partition function ${\cal Z}$ of this theory at the leading order
in a semiclassical approximation (tree-level). We proceed in the following way.
We first integrate over the bulk fields, $\Phi$, constrained to the UV boundary value
$\Phi(x,z=L_0)=\Phi^0(x)$:
\begin{equation}
{\cal Z}[\Phi^0]=\int_{\Phi^0}  d\Phi\ e^{iS(\Phi)}=e^{iS_{\rm eff}(\Phi^0)}\, .
\label{zfunction}
\end{equation}
This is  done simply by obtaining $\Phi$ from their 5D equation
of motions and substituting them back into the action.
The boundary conditions of $\Phi$ at the IR boundary must be chosen
consistently with the 5D variational principle.
For example, for a scalar field with no boundary terms at $z=L_1$,
consistency is guaranteed by taking either a Neumann, $\partial_z\Phi(x,z=L_1)=0$, or a 
Dirichlet, $\Phi(x,z=L_1)=0$, boundary condition. 
The resulting effective action $S_{\rm eff}$ is in general a 4D non-local
action of the UV boundary fields $\Phi^0$.
As a second step, we must integrate over all possible 4D field configurations~$\Phi^0$:
\begin{equation}
{\cal Z}=\int
 d\Phi^0\ e^{i[S_{\rm eff}(\Phi^0)+S_{\rm UV}(\Phi^0)]}\, ,
\end{equation}
where $S_{\rm UV}$ contains any other extra term that $\Phi^0$ 
can have  on the UV boundary.

The 4D boundary action obtained in Eq.~(\ref{zfunction}) 
allows one to establish, at the qualitative level,
the  following ``holographic correspondence'':
the functional ${\cal Z}[\Phi^0]$  in Eq.~(\ref{zfunction}) 
is equivalent to the generating functional obtained
by integrating out a 4D strongly coupled field theory (SCFT)  
in the limit of large number of ``colors'' $N$:
\begin{equation}
{\cal Z}[\Phi^0]=\int d\Phi_{\rm SCFT} \, 
 e^{i[S_{\rm SCFT}+\Phi^0 {\cal O} ]}\, .
\label{adscft}
\end{equation}
Here the fields $\Phi^0$ correspond to external fields 
coupled to the strong sector through  operators ${\cal O}$ made of SCFT fields.
They act like  ``sources'' for correlators of the CFT operators ${\cal O}$.
This correspondence implies that at the classical level the 5D bulk
is equivalent to a 4D SCFT in the large-$N$ limit.

In string theory the holographic correspondence
has been conjectured to be an exact duality
for certain warped geometries~\cite{Maldacena:1997re}.
At the field theoretical level we are considering here, however,
this 5D/4D correspondence is simply based on the observation that 
$n$-point functions defined as 
\be
\langle {\cal O}\cdots {\cal O}\rangle \equiv
\frac{\delta^n \ln{\cal Z}}{\delta \Phi^{0}\cdots\delta \Phi^{0}}\, ,
\label{correlator}
\ee
can be written, both in the 5D theory and in the large-$N$ SCFT, as sums over 
infinitely narrow states. For a 4D strongly coupled theory, 
this decomposition directly follows from the large-$N$ limit~\cite{largeN}.
From the 5D point of view, on the other hand, 
$\langle {\cal O}\cdots {\cal O}\rangle$ 
is computed in terms of 5D propagators, and these can be decomposed as an
infinite sum over 4D propagators of KK modes.
Then, the $n$-point function $\langle {\cal O}\cdots {\cal O}\rangle$
has a similar decomposition as in large-$N$ SCFT.
For example, the two-point function can be written as 
\be
\langle {\cal O}(p){\cal O}(-p)\rangle 
=\sum_{i=1}^{\infty}\frac{F^2_i}{p^2+m^2_i}\, .
\label{2pf}
\ee
In the 4D theory one has $F_i\propto \sqrt{N}$, while in the 5D theory
$F_i\propto 1/g_5$, where $g_5$  parametrizes the loop expansion
($1/g^2_5$ is the coefficient that appears in front of the 5D action).  
For the three-point function, we have
\be
\langle {\cal O}(p_1){\cal O}(p_2){\cal O}(p_3)\rangle 
 = \sum_{i,j,k=1}^{\infty}\lambda_{ijk}(p_1,p_2,p_3)\,
   \frac{F_i}{p_1^2+m^2_i}\, \frac{F_j}{p_2^2+m^2_j}\, \frac{F_k}{p_3^2+m^2_k} \, ,
\label{3pf}
\ee
where $\lambda_{ijk}\propto 1/\sqrt{N}$ in 4D and
$\lambda_{ijk} \propto g_5$ in the 5D theory.

In general, however, we cannot say much 
about the field content of the SCFT, nor about the nature of the operators 
${\cal O}$ that couple to the external fields $\Phi^0$. 
In fact, it is not at all guaranteed that a 4D CFT exists, 
which leads to the same ${\cal Z}[\Phi^0]$ as that of the 5D theory.
Therefore, at the field theoretical level, 
the holographic correspondence should be rather considered 
as a holographic interpretation: 
a qualitative 4D description of a five-dimensional effective field theory.
This interpretation, however, is  very useful to have a clear and quick
qualitative understanding of higher-dimensional theories,
and it is important to develop a  ``dictionary''
that relates the two theories and their properties.
For example, a simple entry in the dictionary is the following: 
 local symmetries of the 5D bulk 
that are broken on the IR boundary
correspond to global symmetries of the 4D SCFT
broken at low-energies $E\sim m_1$;
symmetries of the UV boundary correspond to symmetries
of the external sector $\Phi^0$.
This entry of the dictionary is easy to derive, since any local symmetry
of the 5D bulk is reflected as an  invariance of the effective action
 $S_{\rm eff}$, but not of $S_{\rm UV}$.
Hence,  the  $n$-point functions of Eq.~(\ref{correlator}) 
will fulfill the Ward-Takahashi identities of the symmetry. 
The symmetry breaking  on the IR boundary  affects 
the  $n$-point functions but only
at  distances  $\Delta x$ larger than the 
conformal distance between the two boundaries 
$\Delta x\gtrsim L_1-L_0\sim 1/m_1$.

More can be elaborated on the 5D/4D dictionary when 
the 5D spacetime is Anti deSitter (AdS).
AdS corresponds to 
\begin{equation}
a(z)=L/z\, ,
\end{equation}
where  $L$ is the  AdS curvature radius.
In this case, in the decompactified limit $L_0\rightarrow 0$, $L_1\rightarrow \infty$,
the boundary action $S_{\rm eff}$ is invariant under conformal transformations
due to the AdS isometries~\cite{Maldacena:1997re}.
This implies that, in such a limit, 
the 4D holographic theory is a conformal field theory (CFT), 
and the  operators ${\cal O}$ can be organized according to their dimension.
The momentum scaling of the correlators $\langle {\cal O}\cdots {\cal O}\rangle$ 
is now determined, and this allows us
to derive many properties of the low-energy theory based only on dimensional grounds.
In this case holography becomes a very useful tool to 
understand  5D AdS models.

\section{The effective boundary action for fermions} 
\label{holoforfermions}

Let us compute the boundary action for a theory of a 5D bulk 
fermion~\cite{Henningson:1998cd,Muck:1998iz,PBterm}.
For simplicity, we consider the theory at the quadratic level
(free theory), and derive only  2-point functions. This will be
enough to obtain information about the 4D  spectrum
and derive the holographic interpretation.
A 5D fermion consists in a Dirac field $\Psi=\Psi_L+\Psi_R$, where 
$\gamma_5\Psi_{L,R}=\mp\Psi_{L,R}$. The 5D free action is given by
\begin{equation}
S_5=    \frac{1}{g^2_5}
 \int\! d^4x \int^{L_1}_{L_0}\! dz\, \sqrt{g}
 \left[ \frac{i}{2}\, \bar{\Psi} e^M_A \Gamma^A D_M \Psi
      - \frac{i}{2}\, (D_M \Psi)^\dagger \Gamma^0 e^M_A \Gamma^A \Psi 
      - M\bar{\Psi}\Psi \right],
\label{5dfermion}
\end{equation}
where $\Gamma^A=\{ \gamma^\mu,-i\gamma^5 \}$ denotes the 5D Dirac matrices,
$e^M_A=\delta^M_A/a(z)$ is the inverse vielbein, and
$D_M = \partial_M + \frac{1}{8}\, \omega_{M\, AB} \left[ \Gamma^A, \Gamma^B \right]$ 
the covariant derivative. The only non-vanishing entries in the spin connection 
$\omega_{M\, AB}$ are $\omega_{\mu\, a5}= \eta_{\mu a}/a(z)\, \partial_z a(z)$.
In Eq.~(\ref{5dfermion}) a coefficient $1/g^2_5$ 
has been factored out in front of the action,
so that $g_5$ is the 5D  expansion parameter.
We will work with  Dim$[\Psi]=$3/2, hence Dim$[1/g^2_5]=1$. 

Since the 5D lagrangian for a fermion contains only first-order derivatives,
 one cannot fix simultaneously $\Psi_L$ and $\Psi_R$ on the UV boundary, 
but only one of the two.~\footnote{The Dirac equation relates $\Psi_L$ with $\Psi_R$ 
on the 4D boundaries and therefore they are not independent.}
Let us take as fixed variable -- our ``source'' field -- the left-handed component
\be
\Psi_L(x,z=L_0)=\Psi^0_L\ ,\ \ \ \ \ 
 (\delta\Psi_L=0\ {\rm  on\ the\ UV\ boundary})\, ,
\label{choice}
\ee
and let  $\Psi_R$  be free. 
The variational principle leads to 
\begin{equation} \label{variation}
\begin{split}
0=\delta S_5 = &\frac{1}{g^2_5}
\int\! d^4x \int_{L_0}^{L_1}\! dz\, \sqrt{g}\, 
 \Big[\delta\bar\Psi\, {\cal D}\Psi+ \overline{{\cal D} \Psi}\,
\delta \Psi \Big] \\
 &+ \frac{1}{2}\, \frac{1}{g^2_5} \int\! d^4x\, \sqrt{-g_{\rm ind}}\, 
   \left( \bar\Psi_L\, \delta\Psi_R + \delta\bar\Psi_R\, \Psi_L 
         -\bar\Psi_R\, \delta\Psi_L - \delta\bar\Psi_L\, \Psi_R \right) \Big|_{L_0}^{L_1} \, ,
\end{split}
\end{equation}
where ${\cal D}$ is the 5D Dirac operator. 
The bulk term in the first line of Eq.~(\ref{variation}) vanishes if  the 
5D Dirac equation holds.
The IR boundary term will also vanish if we impose 
a Dirichlet condition $\Psi_L=0$ (or $\Psi_R=0$) on the IR boundary. 
However, the UV boundary term does not vanish, since $\Psi_R$
is, as we said, free to vary. Hence, we must add an extra term to the action
\be
S_{\rm 4}=\frac{1}{2}\, \frac{1}{g^2_5}
 \int_{\rm UV}\! d^4x\, \sqrt{-g_{\rm ind}}\, 
 \left( \bar\Psi_L\Psi_R + \bar\Psi_R\Psi_L  \right) \, ,
\label{extrabound}
\ee
whose variation exactly cancels the UV term in 
Eq.~(\ref{variation}).~\footnote{In Refs.~\cite{Henningson:1998cd,Muck:1998iz} a boundary 
action was introduced by hand in order to derive the AdS/CFT correspondence for a 
theory of fermions. In our approach the boundary term (\ref{extrabound}) is \textit{required} by 
the variational principle, and its coefficient is fixed \cite{PBterm}.}
Extra  terms  on the UV boundary can also be added into the action
if they are only functions of $\Psi_L$ (since on the UV boundary $\delta\Psi_L=0$):
\be
S_{\rm UV}=
\int_{\rm UV}\! d^4x\, \sqrt{-g_{\rm ind}}\
{\cal L}_\text{UV}(\Psi_L)=
\int_{\rm UV}\! d^4x\, \sqrt{-g_{\rm ind}}\
\Big[ \bar\Psi_L i\!\parsl\Psi_L + \dots\Big] \, ,
\label{kinbound}
\ee
where $\parsl\equiv e^\mu_a \Gamma^a \partial_\mu$.
Armed with Eq.~(\ref{5dfermion}) and Eq.~(\ref{extrabound})
we can compute the effective boundary action by integrating out the bulk.
We must  first solve the 5D Dirac equation for the fermion fields,
and then insert this solution back into the action.
Notice that only $S_{\rm 4}$ can give a nonzero contribution,
since the bulk action $S_5$ cancels out on shell.
We look for a solution in momentum space of the form
\be
\Psi_{L,R}(p,z)=\frac{f_{L,R}(p, z)}{f_{L,R}(p, L_0)}\Psi^0_{L,R}(p)\, ,
\label{sollr}
\ee
where $p=\sqrt{p^2}$, and we demand $f_{L,R}(p,z)$ to be solutions of the equations
\begin{equation}  \label{dirac5d}
\left[\partial_z+ 2\,\frac{\partial_z a(z)}{a(z)} \pm a(z)M\right] f_{L,R} =\pm p\, f_{R,L}\, .
\end{equation}
Then, the Dirac equations are satisfied if the 4D field $\Psi^0_R$ 
is related to $\Psi^0_L$ by
\be
 \psl \Psi^0_R= p\frac{f_R(p,L_0)}{f_L(p,L_0)} \Psi^0_L\, . 
\label{dirac4d}
\ee
Inserting Eq.~(\ref{sollr}) and Eq.~(\ref{dirac4d}) into the action,
and rescaling $\Psi_L^0\rightarrow \Psi_L^0/\sqrt{a(L_0)^3}$
such that the kinetic term on the UV boundary lagrangian~(\ref{kinbound}) 
is canonically normalized,
we have that the whole boundary action is given by
\be
S_\text{UV} + \int\! \frac{d^4p}{(2\pi)^4}\, \bar\Psi^0_L\Sigma(p)
\Psi^0_L\, ,\ \ \ \ \ {\rm  where}\ \ \ \ \ 
\Sigma(p)=
\frac{ a(L_0)}{g^2_5}\,\frac{p}{\psl}\,\frac{f_R(p,L_0)}{f_L(p,L_0)}\, .
\label{effbound}
\ee
The function $\Sigma(p)$ corresponds to the correlator $\langle{\cal OO}\rangle$
of the 4D holographic theory. Its poles will determine  the 4D mass spectrum. 

At this stage it can be useful to make contact with the usual Kaluza-Klein
description of theories in a slice of 5D spacetime or orbifolds.
In the Kaluza-Klein approach, the cancellation of Eq.~(\ref{variation}) 
is accomplished by imposing a Dirichlet condition on the UV boundary either
for the right- or for the left-handed component of the fermion field.
For example,  ignoring possible UV boundary terms, this condition is either 
$\Psi_R=0$ or $\Psi_L=0$. These two cases correspond in orbifold theories
respectively to even ($+$) and odd ($-$) parity for the bulk fermion field $\Psi_L$.
In our approach the value of the field on the UV boundary is determined by the 
equations of motion for $\Psi_L^0$ after the bulk has been integrated out.
For example, our choice Eq.~(\ref{choice}) corresponds, on the orbifold, 
to an even $\Psi^0_L$. Indeed, the 4D equation of motion for $\Psi^0_L$ 
(as derived from Eqs.~(\ref{extrabound}) and (\ref{kinbound})) 
gives exactly the same  boundary condition  for a $\Psi_L$ even
in the orbifold
({\it e.g.} $\Psi^0_R=0$ if $S_\text{UV}=0$, that is, if no extra 
UV boundary terms are introduced).
To reproduce in our formalism an odd UV boundary condition for $\Psi_L$,
we have to add an extra fermion $\Psi^\prime_R$ on the UV boundary 
(a Lagrange multiplier) with  the lagrangian
\be
{\cal L}_\text{UV}= \frac{a(L_0)}{g^2_5}\bar\Psi^\prime_R\Psi^0_L+h.c.
+{\cal L}(\Psi^\prime_R)\, ,
\label{odd}
\ee
where ${\cal L}(\Psi^\prime_R)$ includes possible extra terms for $\Psi^\prime_R$ 
(the same as those for the even field $\Psi_R^0$ on the UV boundary of an orbifold: 
${\cal L}(\Psi_R^\prime)={\cal L}_{\rm orb}(\Psi_R^0\rightarrow\Psi_R^\prime)$). 
One can easily check that the equations of motion of $\Psi^\prime_R$ and $\Psi^0_L$
 lead to the orbifold boundary condition for an odd $\Psi_L$.
For example, for the case ${\cal L}(\Psi^\prime_R)=0$ we obtain $\Psi_L^0=0$.
In this case $\Psi_L^0$ is frozen and  acts like a classical source coupled to the CFT.

\section{The holographic description of AdS$_5$ with two boundaries}
\label{AdSholo}

We specialize here to the AdS case, whose holographic description 
is particularly interesting.~\footnote{For the holographic interpretation
in the case of bosonic fields in AdS$_5$ with boundaries, 
see~\cite{arpora,Rattazzi:2000hs,Perez-Victoria:2001pa}.}
Using the solution for $f_{L,R}$ given in the Appendix, we obtain
\begin{equation} \label{sigma}
\Sigma(p)= \frac{L}{g^2_5L_0}\,\frac{p}{\psl}\, 
 \frac{J_{\alpha-1}(pL_0) Y_{\beta}(pL_1)- J_{\beta}(pL_1) Y_{\alpha-1}(pL_0)}%
      {J_{\alpha}(pL_0)   Y_{\beta}(pL_1)- J_{\beta}(pL_1) Y_{\alpha}(pL_0)}\, ,
\end{equation}
where $p=\sqrt{p\cdot p}$ and
\begin{equation}
\alpha = ML + \frac{1}{2} \, ; \qquad\quad
 \beta = \begin{cases} \alpha-1 & \quad \text{for L$_+$} \\ 
                       \alpha   & \quad \text{for L$_-$}\, . \end{cases}
\end{equation}
Here L$_-$ (L$_+$) is a shorthand notation for the case 
of Dirichlet boundary condition $\Psi_L=0$ ($\Psi_R=0$) 
on the IR boundary. Depending on the value of $M$, we have different  
holographic interpretations of this theory.

\subsection{The case  $ML \geq 1/2$}  

Let us first present the holographic description and show later that it indeed agrees
with the 5D result.
The 4D theory consists of an external dynamical field $\Psi_L^0$
coupled to a strongly interacting CFT. 
The holographic lagrangian is 
\be
{\cal L}={\cal L}_{\rm CFT}+{\cal L}_\text{UV}(\Psi^0_L)
 + Z_0\, \bar\Psi^0_L i\!\parsl \Psi^0_L
 + \frac{\omega}{\Lambda^{\alpha-1}} \left(\bar\Psi^0_L{\cal O}_R+h.c.\right)
 + \xi\, \frac{\bar{\cal  O}_R\parsl{\cal O}_R}{\Lambda^{2\alpha}}+ \dots \, ,
\label{holo}
\ee
where ${\cal O}_R$ is a CFT chiral operator of dimension
\be
{\rm Dim}[{\cal O}_R] = \frac{3}{2} + \left| ML + \frac{1}{2} \right|\, ,
\label{dime}
\ee
and $Z_0$, $\omega$, $\xi$ are dimensionless running couplings.
Due to the UV boundary, the conformal symmetry
is broken at energies above  the scale $\Lambda=1/L_0$. 
In Eq.~(\ref{holo}) we have considered the effective theory below $\Lambda$.
CFT deformations then arise from higher-dimensional terms suppressed by $\Lambda$, 
and in Eq.~(\ref{holo}) we only show the dominant one.
Since we are assuming $ML\geq 1/2$, (so that $\alpha \geq 1$), the coupling
of the elementary fermion $\Psi^0_L$ to the CFT is always irrelevant, at most 
marginal if $ML=1/2$ ($\alpha =1)$ -- in this latter 
case the coupling has  a logarithmic dependence upon $\Lambda$.
The presence of the IR boundary in the 5D theory corresponds in 4D to an IR cutoff at 
the scale $\mu=1/L_1$. This means that there is a mass gap in the theory and 
the CFT spectrum  is  discretized. 
Bound states have masses of order $m_{\rm CFT}\sim n\pi \mu$ ($n=1,2,\dots$).

If the lagrangian (\ref{holo}) gives
the correct holographic interpretation, this must lead,
 after integrating out the CFT at leading
order in the number of ``colors'' $N$, 
to Eq.~(\ref{effbound})
with $\Sigma(p)$ given by Eq.~(\ref{sigma}). 
Of course, this is not possible without having defined the exact CFT,
and even if the latter is known (as it can be the case in certain 5D supergravity models),
we do not know how to compute $\Sigma(p)$ in the strongly coupled regime.
Nevertheless, one can show that the properties of $\Sigma(p)$ which can be deduced 
from the 4D theory defined by Eq.~(\ref{holo}) are also properties of 
Eq.~(\ref{sigma}). 
These 4D properties are:

\noindent \textsc{I. Scale invariance:}
Performing the rescaling  $\Psi^0_L\rightarrow \Lambda^{\alpha-1} \Psi^0_L$ and
taking the limit $\Lambda\rightarrow \infty$ 
and $\mu\rightarrow 0$ in Eq.~(\ref{holo}),
we obtain that the kinetic term of 
$\Psi^0_L$  tends to infinity and $\Psi^0_L$ becomes
an external frozen source probing the CFT:
\be
{\cal L}={\cal L}_{\rm CFT}+ \omega \left(\bar\Psi^0_L{\cal O}_R+h.c.\right)\, .
\ee
Scale invariance then dictates the $p$-dependence of 
the correlator function:
\be
\langle {\cal O}_R(p) \bar {\cal O}_R(-p) \rangle
\propto \ \psl\ p^{(2 {\rm Dim}[{\cal O}_R]-5)}=
\ \psl\ p^{(2 \alpha-2)}\, .
\label{scaling}
\ee
The correlator $\langle {\cal O}_R \bar {\cal O}_R\rangle$ is given by
$\Sigma(p)$ once local terms which diverge in the limit $\Lambda\to\infty$, 
like the kinetic term, are removed by adding appropriate counterterms.

\noindent \textsc{II. Large N:}
Using a large-$N$ expansion, we can infer some properties 
of $\langle {\cal O}_R \bar {\cal O}_R\rangle$.  
Indeed, similarly to the case of QCD at large $N$, where current correlators can be 
decomposed as the infinite exchange of colorless mesons~\cite{largeN},
in our case too the two-point function 
$\langle {\cal O}_R \bar {\cal O}_R\rangle$ can be written, in the large-$N$ limit,
as a sum over ``colorless'' fermionic states:
\be
\langle {\cal O}_R(p) \bar {\cal O}_R(-p) \rangle
=\sum_{i=1}^{\infty}\frac{F^2_i}{p^2+m^2_i}\, .
\label{sum}
\ee

\noindent \textsc{III. UV cutoff effects:}
The presence of a UV cutoff affects the theory in two ways.
First, higher-dimensional operators suppressed by $1/\Lambda$ are present in the theory. 
These operators modify the mass spectrum. For example, the most important one, 
$(\bar{\cal  O}_R\parsl{\cal O}_R)/\Lambda^{2\alpha}$,
induces a correction of order
\be
\Delta m_\text{CFT} \sim 
 \left( \frac{m_\text{CFT}}{\Lambda} \right)^{2\alpha}\, m_\text{CFT} \, .
\label{shift}
\ee
A second effect of the UV cutoff is that the field $\Psi_L^0$   can  propagate
and mix with the CFT bound states. 
The modified spectrum is given by
the poles of the whole inverse quadratic term in $S_\text{eff}+S_{\rm UV}$.
In absence of the extra boundary action $S_\text{UV}$, 
the spectrum is then determined by the zeros of $\Sigma(p)$. 
If the CFT does not contain any massless state (it is not chiral),
the massless eigenstate will be mostly $\Psi^0_L$
since the mixing of the latter with the CFT is always irrelevant or, at most, marginal.

Let us turn to the 5D result. We want to show that the above properties
are also satisfied by Eq.~(\ref{sigma}). 
Rotating Eq.~(\ref{sigma}) to Euclidean momenta, and
taking the limit $\Lambda=1/L_0\to\infty$, $\mu=1/L_1\rightarrow 0$ 
one obtains
\begin{gather} 
\begin{split} \label{Sigma5Dexp}
\Sigma(p) \simeq i\psl \, \frac{L}{g^2_5}\,
 \bigg\{ &a_1+a_2\; (pL_0)^2 + \dots \\
 &+b_1\; (pL_0)^{(2\alpha-2)}+ \dots \bigg\} \, ,
\end{split}  \\[0.3cm]
a_1 = \frac{1}{2(\alpha-1)}\, ,\qquad\quad a_2 = -\frac{1}{8(\alpha-2)(\alpha-1)^2} \, ,\qquad\quad 
 b_1 = -2^{-2\alpha+1}\,\frac{\Gamma(1-\alpha)}{\Gamma(\alpha)}\, ,
\end{gather}
where only
the first two analytic terms in an expansion in $p L_0$ 
(first line of Eq.~(\ref{Sigma5Dexp})) and the first among the non-analytic terms 
(second line) are shown.~\footnote{In the particular case of 
integer $\alpha$ the expansion 
is different, since the first non-analytic term develops a $\log p$.}
The first analytic term corresponds to the   ``bare'' kinetic term of $\Psi^0_L$
in Eq.~(\ref{holo});  thus we can match them  at high energy:
$Z_0(\Lambda)=a_1\, L/g_5^2$. 
The two-point correlator $\langle {\cal O}_R\bar {\cal O}_R\rangle$ corresponds 
to the first non-analytic term in the expansion, 
whose form agrees with Eq.~(\ref{scaling}).
It can be extracted by rescaling $\Psi^0_L\to L_0^{1-\alpha}\Psi^0_L$
and taking the limit $L_0\to 0$. Local terms which diverge 
in this limit can always be canceled by suitable counterterms in 
$S_{\rm UV}$.~\footnote{This is the standard ``renormalization'' procedure
adopted to derive the AdS/CFT correspondence in string theory.}
Thus
\begin{equation} \label{AdSCFT}
\langle {\cal O}_R(p) \bar{{\cal O}}_R(-p) \rangle = 
 \lim_{L_0\to 0} 
 \left( \Sigma(p) + \text{counterterms} \right)\, .
\end{equation}
The poles of  $\langle {\cal O}_R(p) \bar{{\cal O}}_R(-p) \rangle$
and $\Sigma(p)$ coincide, since they only differ by local terms.

Property (\ref{sum}) can then be easily verified.
Since $f_{L,R}$ are analytic functions with only simple zeros 
on the real axis, $\Sigma(p)$  has only simple poles 
and can be expanded as in Eq.~(\ref{sum}). 
The poles of $\Sigma(p)$ correspond to the zeros of $f_{L}$:
\footnote{This is true at least for the massive poles. To identify possible 
massless poles, that is poles at $p^2=0$, one must look at the whole $\Sigma(p)$,
not just at its denominator.}
\begin{equation} \label{zeros}
f_L(p,L_0) = 0 \quad \Leftrightarrow \quad 
 J_{\alpha}(p L_0) Y_{\beta}(p L_1) - J_{\beta}(p L_1) Y_{\alpha}(p L_0)=0\, .
\end{equation}
The masses $m_i$ of the CFT spectrum are then obtained 
by taking the limit $\Lambda\rightarrow\infty$:
\begin{equation} \label{zeromodeh}
 f_L(m_i,L_0) = 0 \quad \Leftrightarrow \quad J_{\beta}(m_i L_1)=0\, .
\end{equation}

The first effect of a finite UV cutoff is that higher-dimensional operators
deform the unperturbed CFT spectrum (\ref{zeromodeh}) to that of Eq.~(\ref{zeros}).
Expanding~(\ref{zeros}) for large but finite $\Lambda$, 
and using~(\ref{zeromodeh}), one can estimate the level distortion $\Delta m_i$:
\begin{equation}
\Delta m_i  \simeq 
\frac{ c}{L_1} 
 \frac{Y_\beta(m_i L_1)}
{J_\beta^\prime(m_i L_1)}
\left(\frac{m_i}{\Lambda}\right)^{2\alpha}
\,  ,
\end{equation}
where $c$ is a coefficient which depends on $\alpha$.
This result is in agreement with our expectation Eq.~(\ref{shift}).
The second effect of the UV cutoff is the mixing between $\Psi_L^0$ and the CFT
bound states. As explained before, 
neglecting UV boundary terms ($S_{\rm UV}$),
the modified spectrum is now given by the \textit{zeros} of $\Sigma(p)$, 
instead of by its poles:
\begin{equation} \label{spectrumfinal} 
f_R(p,L_0) = 0 \quad \Leftrightarrow \quad 
 J_{\alpha-1}(p L_0) Y_{\beta}(p L_1) - J_{\beta}(p L_1) Y_{\alpha-1}(p L_0)=0\, .
\end{equation}

We should notice at this point that Eqs.~(\ref{zeros}) and (\ref{spectrumfinal}) 
precisely correspond to the spectra of massive KK levels for an orbifold theory 
respectively with UV parities $(-)$ and $(+)$
for $\Psi_L$ (and opposite parities for $\Psi_R$). 
This is not a surprise, since we already argued
at the end of section~\ref{holoforfermions} that an orbifold theory with $\Psi_L$ odd 
on the UV boundary does correspond in our approach to the case 
in which $\Psi_L^0$ is 
frozen and  acts like a classical source coupled to the CFT.
An orbifold theory with $\Psi_L$ even, on the other hand,
corresponds, in our formalism, to the case in which $\Psi_L^0$ is dynamical and modifies
the CFT spectrum. We are thus explicitly verifying that the physical spectrum is
the same, as it should be, both in our holographic description and in the Kaluza-Klein 
description.

This equivalence also gives us information about the presence of massless states 
in the CFT spectrum. Indeed, we know that in orbifold compactifications
if $\Psi_L$ has parities $(-,-)$  under reflections with respect to the two
boundaries, then $\Psi_R$ is $(+,+)$ and the spectrum contains
a right-handed massless mode.
This implies that when a Dirichlet boundary condition $\Psi_L=0$ 
is imposed on the IR boundary, (the L$_-$ case),
the CFT must be chiral and form a massless bound state.
This can be easily checked, since a massless state of the CFT appears 
like a pole in $\Sigma(p)$ for $p\to 0$. 
Expanding Eq.~(\ref{sigma}) for $p\ll 1/L_1$ with $\beta=\alpha$, one finds
\begin{equation}
\Sigma(p) \simeq -2\alpha \frac{L}{g_5^2}\, \frac{1}{\psl}\, \frac{1}{L_1^2}  
 \left(\frac{L_0}{L_1}\right)^{2(\alpha-1)} + \dots
\end{equation}
thus confirming our expectation. 
The factor $\left(L_0/L_1\right)^{2(\alpha-1)}$ correctly appears, as the
massless right-handed bound state of the CFT is excited by the external left-handed
source through the coupling of Eq.~(\ref{holo}).
No massless pole exists in the case L$_+$ for $ML>1/2$.
A pictorial representation of the holographic theory is given in Fig.~\ref{fig:Lpm}.
 \FIGURE[t]{
 \centering \small \hspace{55pt}
    \begin{picture}(300,100)
      \SetWidth{1} \LongArrow(10,75)(260,75) \SetWidth{0.5}
      \Line(120,72)(120,78) \Line(180,72)(180,78) \Text(-15,30)[rc]{\large {\bf L$_+$}:}
      \Text(120,80)[cb]{$-1/2$} \Text(180,80)[cb]{$+1/2$} \Text(10,80)[lb]{$ML$}
      \Line(119,67)(119,7) \Line(121,67)(121,7)
      \SetWidth{0.8} \CArc(50,30)(30,0,360) \SetWidth{0.4} 
      \Line(36,30)(64,30) \Line(36,33)(64,33) \Line(36,36)(64,36) 
      \Line(36,38)(64,38) \Line(36,40)(64,40) \Line(36,41.5)(64,41.5) 
      \Line(36,17)(64,17) \Text(66,17)[lc]{\scriptsize $L$} \SetWidth{0.5} 
      \SetColor{BrickRed} 
      \Vertex(70.8,43.9){1} \Line(70.8,43.9)(87.4,55.0) \Text(89.1,56.1)[cl]{\BrickRed{$\Psi_L^0$}}
      \SetColor{Black}
      \Vertex(74.5,34.9){1} \Line(74.5,34.9)(94.1,38.8) \Text(96.1,39.2)[cl]{$\chiR$}
      \SetWidth{0.8} \CArc(200,30)(30,0,360) \SetWidth{0.4} 
      \Line(186,30)(214,30) \Line(186,33)(214,33) \Line(186,36)(214,36) 
      \Line(186,38)(214,38) \Line(186,40)(214,40) \Line(186,41.5)(214,41.5) 
      \SetColor{BrickRed} 
      \Vertex(220.8,43.9){1} \Line(220.8,43.9)(237.4,55.0) \Text(239.1,56.1)[cl]{\BrickRed{$\Psi_L^0$}}
      \SetColor{Black}
    \end{picture} \vspace*{0.5cm} \newline  \hspace*{40pt}
    \begin{picture}(300,100)
      \SetWidth{1} \LongArrow(10,75)(260,75) \SetWidth{0.5}
      \Line(120,72)(120,78) \Line(180,72)(180,78) \Text(-15,30)[rc]{\large {\bf L$_-$}:}
      \Text(120,80)[cb]{$-1/2$} \Text(180,80)[cb]{$+1/2$} \Text(10,80)[lb]{$ML$}
      \Line(119,67)(119,7) \Line(121,67)(121,7)
      \SetWidth{0.8} \CArc(50,30)(30,0,360) \SetWidth{0.4} 
      \Line(36,30)(64,30) \Line(36,33)(64,33) \Line(36,36)(64,36) 
      \Line(36,38)(64,38) \Line(36,40)(64,40) \Line(36,41.5)(64,41.5) 
      \SetColor{BrickRed}
      \Vertex(70.8,43.9){1} \Line(70.8,43.9)(87.4,55.0) \Text(89.1,56.1)[cl]{\BrickRed{$\Psi_L^0$}}
      \SetColor{Black}
      \Vertex(74.5,34.9){1} \Line(74.5,34.9)(94.1,38.8) \Text(96.1,39.2)[cl]{$\chiR$}
      \SetWidth{0.8} \CArc(200,30)(30,0,360) \SetWidth{0.4} 
      \Line(186,30)(214,30) \Line(186,33)(214,33) \Line(186,36)(214,36) 
      \Line(186,38)(214,38) \Line(186,40)(214,40) \Line(186,41.5)(214,41.5) 
      \Line(186,17)(214,17) \Text(216,17)[lc]{\scriptsize $R$} \SetWidth{0.5} 
      \SetColor{BrickRed}
      \Vertex(220.8,43.9){1} \Line(220.8,43.9)(237.4,55.0) \Text(239.1,56.1)[cl]{\BrickRed{$\Psi_L^0$}}
      \SetColor{Black}
    \end{picture}
 \caption{\it Holographic theories for the cases L$_+$ and L$_-$, schematically
 represented as a CFT sector coupled to an external source $\Psi_L^0$ (in red) and
 to an elementary field $\chiR$ (for $ML<-1/2$). When both present, $\Psi_L^0$ and $\chiR$ 
 marry through a mass mixing term and become heavy. The CFT spectrum is represented by 
 horizontal lines; an isolated line indicates a massless chiral bound state. 
 When this state is present, it marries with $\Psi_L^0$ or $\chiR$.}
 \label{fig:Lpm}
 }

\subsection{The case $-1/2 \leq ML \leq 1/2$}

For these values of the 5D fermion mass, the holographic theory
is the same as that of Eq.~(\ref{holo}) and 
(\ref{dime}).
Most of the results obtained in the previous section, for example what we argued
about the spectrum, apply to this case as well. 
However, there is an important difference. We have  now that
Dim$[{\cal O}_R]\leq 5/2$
($0\leq \alpha\leq 1$), and this means that the coupling of the external field
$\Psi^0_L$ to the CFT is relevant
at low energies.
In other words, the effect of the mixing of $\Psi^0_L$
with the CFT states  is always important, and
the external field does not decouple even 
in the limit $\Lambda\rightarrow\infty$.
The CFT spectrum suffers modifications of order one  
and the massless eigenstate becomes a mixture of $\Psi^0_L$ and the CFT.
Notice that, contrary to Ref.~\cite{Muck:1998iz},
we find that we can still choose a left-handed source even if  $M<0$.
The boundary action is still well-defined and 
the 5D/4D correspondence works fine.

\subsection{The case $ML \leq -1/2$}

For $ML \leq -1/2$ ($\alpha<0$) there is still a sensible holographic
description of our theory, although quite different from the previous cases,
as we now show.
Rotating Eq.~(\ref{sigma}) to Euclidean momenta, and expanding for $1/L_0 \to\infty$, 
$1/L_1\to 0$, now gives
\begin{gather} 
\begin{split}
\Sigma(p) \simeq i\psl \, \frac{L}{g^2_5}\,
 \bigg\{ &\frac{\tilde a_0}{(pL_0)^2} + \tilde a_1 +\tilde a_2\; (pL_0)^2 + \dots \\
 &+\tilde b_1\; (pL_0)^{(2|\alpha|-2)}+ \dots \bigg\} \, ,
\end{split} \label{Sigma5Dexp2} \\[0.3cm]
\tilde a_0 =2|\alpha| \, ,\qquad\quad \tilde a_1 = \frac{1}{2(|\alpha|-1)} \, ,\qquad\quad 
 \tilde b_1 = 2^{2\alpha+1}\,\frac{\Gamma(1+\alpha)}{\Gamma(-\alpha)}\, , \label{coeff2}
\end{gather}
which differs from (\ref{Sigma5Dexp}) for the appearance of a pole at $p^2=0$.
Being the latter a non-analytic term, it cannot be simply canceled by some
local counterterm, with the result that Eq.~(\ref{AdSCFT}) does not hold in this case.
Quite clearly, the pole cannot be ascribed to the exchange of some
massless bound state of the CFT sector, being the conformal symmetry restored 
in the limit $\mu=1/L_1\to 0$.
We conclude that the holographic interpretation of Eq.~(\ref{Sigma5Dexp2}) 
does not involve only a CFT sector coupled to the $\Psi_L^0$ external field, 
but it must include some additional \textit{elementary} degree of freedom, 
whose imprint is the pole term.
This new field is excited by the left-handed source $\Psi^0_L$, and thus couples to it.
The term in the second line of Eq.~(\ref{Sigma5Dexp2})
can  be truly 
interpreted as the two-point CFT correlator $\langle {\cal O}_R \bar {\cal O}_R \rangle$
with the dimension of ${\cal O}_R$
given by Eq.~(\ref{dime}).

The correct picture turns out to be the following: a right-handed elementary field 
$\chiR$ exists, which has a mass mixing with $\Psi^0_L$, and couples to the CFT through
the same operator ${\cal O}_R$ that appears in the coupling of $\Psi^0_L$ to the CFT.
The holographic lagrangian is given by
\begin{equation} \label{holo2}
\begin{split}
{\cal L}=& {\cal L}_{\rm CFT}+{\cal L}_\text{UV}(\Psi^0_L) + Z_0\, \bar\Psi^0_L i\!\parsl\Psi_L^0
 + \tilde Z_0\, \chiRbar i\!\parsl\chiR 
 +  \eta\Lambda \left( \chiRbar \Psi^0_L +\bar\Psi^0_L \chiR \right) \\ 
 &+ \left[\frac{\omega}{\Lambda^{|\alpha|-1}}\, \bar\Psi^0_L{\cal O}_R
 + \frac{\tilde\omega}{\Lambda^{|\alpha|}}\, \chiRbar\parsl{\cal O}_R+ h.c.\right]
 + \xi\,\frac{\bar{\cal  O}_R\parsl{\cal O}_R}{\Lambda^{2|\alpha|}}+\dots\, ,
\end{split}
\end{equation}
where the dimension of the operator ${\cal O}_R$ is still given by Eq.~(\ref{dime}).
From Eq.~(\ref{Sigma5Dexp2}) 
we can extract the high-energy value of some of the couplings;
for instance: $Z_0(\Lambda)= \tilde a_1\, L/g_5^2$, 
$\eta^2/\tilde Z_0 =\tilde a_0\, L/g_5^2$. 
We see that   the new state $\chiR$ marries the external field $\Psi^0_L$
and both become massive. Notice that this mass is of order the
cutoff $\Lambda$.
Hence, the elementary field $\chiR$ plays the same  role as 
$\Psi^\prime_R$ in Eq.~(\ref{odd}): it forces  
the source $\Psi^0_L$ not to propagate at low energies.
The theory is depicted in Fig.~\ref{fig:Lpm}.

To obtain the massive spectrum of the pure CFT from $\Sigma(p)$
we can again take the limit $\Lambda\to\infty$.
In this limit, any distortion from the new elementary field $\chiR$ disappears,
being the coupling of $\chiR$ to the CFT  always irrelevant.
Expanding Eq.~(\ref{zeros}) for $L_0\to 0$ we obtain:
\begin{equation} \label{zeromodeh2}
f_L(m_i,L_0) = 0 \quad \Leftrightarrow \quad  
 Y_\beta(m_i L_1)\tan\alpha\pi - J_{\beta}(m_i L_1)=0\, .
\end{equation}
The effect of a finite cutoff is again that of deforming the CFT spectrum.
Keeping the leading correction and using (\ref{zeromodeh2}), one has
\begin{equation}
 \Delta m_i\simeq \frac{\tilde c}{L_1}
 Y_{\beta}(m_i L_1)
 \left[J_\beta^\prime(m_i L_1)-Y_\beta^\prime(m_i L_1)\tan\alpha\pi \right]^{-1}
\left(\frac{m_i}{\Lambda}\right)^{2|\alpha|}\, ,
\end{equation}
where $\tilde c$ is a coefficient which depends on $\alpha$.
The level distortion
$\Delta m_i\sim m_i (m_i/\Lambda)^{2|\alpha|}$ can be explained as the effect 
of the operator $(\bar{\cal O}_R\parsl {\cal O}_R)/\Lambda^{2|\alpha|}$, though in this case an 
additional correction of the same size comes from the mixing of $\chiR$ with the CFT 
bound states.

The massless spectrum is again quite interesting. 
As before, from the equivalence with the KK spectrum of an orbifold
theory with $\Psi_L(-,-)$, we deduce that 
$\Sigma(p)$ must contain a right-handed massless mode in the case L$_-$.
This time, however, such massless mode corresponds to the elementary
field $\chiR$, rather than to some CFT bound state.
Indeed, expanding $\Sigma(p)$ for $p\ll 1/L_1$ (case L$_-$), one finds
\begin{equation} \label{pole}
\Sigma(p)\simeq 2\alpha \frac{L}{g_5^2}\,\frac{1}{\psl}\,\frac{1}{L_0^2} + \dots
\end{equation}
This pole is evidently generated by the exchange of $\chiR$ through its mass mixing
with the source $\Psi_L^0$. No dependence upon the IR scale $1/L_1$ appears, as one 
expects for a truly external mode.
Hence, the CFT spectrum is \textit{not} chiral in this case.
The opposite happens in the L$_+$ case (see Fig.~\ref{fig:Lpm}): 
we know from the orbifold KK description that if $\Psi_L$ has parities $(-,+)$,
the complete spectrum does not have any massless chiral state. Since we also know
that a $\chiR$ chiral field already exists which could lead to a massless pole,
then we conclude that the CFT must be chiral. That is, a left-handed chiral bound state
must exists which marries $\chiR$, so that no massless mode appears in the final spectrum. 
Our guess can be checked by studying $\Sigma(p)$ in the limit $p\to 0$: expanding
Eq.~(\ref{sigma}) for $p\ll L_1$, one has
\begin{equation}
\Sigma(p)\simeq 2\alpha \frac{L}{g_5^2}\,\psl\,
 \frac{1-(L_0/L_1)^{2\alpha-2}}{4\alpha(\alpha-1)-(pL_1)^2 (L_0/L_1)^{2\alpha}}\, .
\end{equation}
If we fix $L_0$ to a small but finite value, and look at the regime $p\to 0$, then
we see that there is no massless pole. The leading term in the expansion
\begin{equation}
\Sigma(p)\simeq \psl\,\frac{L}{g_5^2} \frac{1}{2(1-\alpha)} 
 \left(\frac{L_0}{L_1}\right)^{2\alpha-2} + \dots \, ,
\end{equation}
is what one would expect from the exchange of a massive $\chiR$, whose tiny mass
$\tilde m\sim \mu (\mu/\Lambda)^{|\alpha|}$ is generated through its coupling to the CFT.
On the other hand, if we first send $L_0$ to zero, thus decoupling the CFT from $\chiR$,
then we \textit{do} find a massless pole, being the expansion for $\Sigma(p)$
the same as in Eq.~(\ref{pole}). This is again expected, since $\chiR$ is massless
once decoupled from the CFT. Thus, the whole picture is consistent with a
chiral CFT spectrum in the case L$_+$.

\subsection{Holography with a Right-handed source}
\label{RHsource}

An alternative holographic description can also be given
in terms of the right-handed source. 
Let us fix the value of $\Psi_R$, instead of $\Psi_L$,
on the UV boundary:
\be
\Psi_R(x,z=L_0)=\Psi^0_R\ ,\ \ \ \ \ 
 (\delta\Psi_R=0\ {\rm  on\ the\ UV\ boundary})\, .
\ee
The holographic description can be easily deduced by noting that 
if we use Eq.~(\ref{dirac4d}) and rewrite the boundary action in terms of $\Psi_R^0$,
the function $\Sigma(p)$ turns out to be the inverse 
(except for an overall coefficient) 
of that for a left source:
\begin{equation}
S_\text{UV} + \int\! \frac{d^4p}{(2\pi)^4}\, \bar\Psi^0_R\Sigma(p) \Psi^0_R\, ,
 \qquad \Sigma(p)= 
 \frac{ a(L_0)}{g^2_5}\,\frac{\psl}{p}\,\frac{f_L(p,L_0)}{f_R(p,L_0)}\, .
\end{equation}
Furthermore, the system of Dirac equations (\ref{dirac5d}) is invariant under 
the symmetry:
\begin{equation} \label{symmetry}
L \leftrightarrow R\ , \qquad M \to - M\, ,
\end{equation}
provided that the relative sign between $f_L$ and $f_R$ is also changed.
This implies that
\begin{equation}
\Sigma_\text{R$_\pm$} = - \Sigma_\text{L$_\pm$} \{ M\to -M\} \qquad \forall\, M \, ,
\end{equation}
having denoted with R$_{-}$ (R$_{+}$) the case where a Dirichlet boundary condition 
$\Psi_R=0$ ($\Psi_L=0$) is imposed on the IR boundary.
In other words, for a given bulk mass $M$, the holographic description in terms of a right-handed 
source is equivalent to the description with a left-handed source at $-M$, with inverted chiralities,
see Fig.~\ref{fig:Rpm}.
In particular, the elementary fields will couple to the CFT sector
through a left-handed chiral operator ${\cal O}_L$ of dimension 
\begin{equation}
\text{Dim}[{\cal O}_L]=\frac{3}{2}+\left|ML-\frac{1}{2}\right|\, ,
\end{equation}
and an extra external left-handed field $\chiL$ appears for $ML>+1/2$. 
\FIGURE[t]{
\centering \small \hspace{14pt}
   \begin{picture}(300,100)
     \SetWidth{1} \LongArrow(40,75)(290,75) \SetWidth{0.5}
     \Line(120,72)(120,78) \Line(180,72)(180,78) \Text(15,30)[rc]{\large {\bf R$_-$}:}
     \Text(120,80)[cb]{$-1/2$} \Text(180,80)[cb]{$+1/2$} \Text(40,80)[lb]{$ML$}
     \Line(179,67)(179,7) \Line(181,67)(181,7)
     \SetWidth{0.8} \CArc(80,30)(30,0,360) \SetWidth{0.4} 
     \Line(66,30)(94,30) \Line(66,33)(94,33) \Line(66,36)(94,36) 
     \Line(66,38)(94,38) \Line(66,40)(94,40) \Line(66,41.5)(94,41.5) 
     \Line(66,17)(94,17) \Text(96,17)[lc]{\scriptsize $L$} \SetWidth{0.5} 
     \SetColor{BrickRed} 
     \Vertex(100.8,43.9){1} \Line(100.8,43.9)(117.4,55.0) \Text(119.1,56.1)[cl]{\BrickRed{$\Psi_R^0$}}
     \SetColor{Black}
     \SetWidth{0.8} \CArc(230,30)(30,0,360) \SetWidth{0.4} 
     \Line(216,30)(244,30) \Line(216,33)(244,33) \Line(216,36)(244,36) 
     \Line(216,38)(244,38) \Line(216,40)(244,40) \Line(216,41.5)(244,41.5) 
     \SetColor{BrickRed} 
     \Vertex(250.8,43.9){1} \Line(250.8,43.9)(267.4,55.0) \Text(269.1,56.1)[cl]{\BrickRed{$\Psi_R^0$}}
     \SetColor{Black}
     \Vertex(254.5,34.9){1} \Line(254.5,34.9)(274.1,38.8) \Text(276.1,39.2)[cl]{$\chiL$}     
   \end{picture} \vspace*{0.5cm} \newline  \hspace*{-20pt}
   \begin{picture}(300,100)
     \SetWidth{1} \LongArrow(40,75)(290,75) \SetWidth{0.5}
     \Line(120,72)(120,78) \Line(180,72)(180,78) \Text(15,30)[rc]{\large {\bf R$_+$}:}
     \Text(120,80)[cb]{$-1/2$} \Text(180,80)[cb]{$+1/2$} \Text(40,80)[lb]{$ML$}
     \Line(179,67)(179,7) \Line(181,67)(181,7)
     \SetWidth{0.8} \CArc(80,30)(30,0,360) \SetWidth{0.4} 
     \Line(66,30)(94,30) \Line(66,33)(94,33) \Line(66,36)(94,36) 
     \Line(66,38)(94,38) \Line(66,40)(94,40) \Line(66,41.5)(94,41.5) 
     \SetColor{BrickRed} 
     \Vertex(100.8,43.9){1} \Line(100.8,43.9)(117.4,55.0) \Text(119.1,56.1)[cl]{\BrickRed{$\Psi_R^0$}}
     \SetColor{Black}
     \SetWidth{0.8} \CArc(230,30)(30,0,360) \SetWidth{0.4} 
     \Line(216,30)(244,30) \Line(216,33)(244,33) \Line(216,36)(244,36) 
     \Line(216,38)(244,38) \Line(216,40)(244,40) \Line(216,41.5)(244,41.5) 
     \Line(216,17)(244,17) \Text(246,17)[lc]{\scriptsize $R$} \SetWidth{0.5} 
     \SetColor{BrickRed} 
     \Vertex(250.8,43.9){1} \Line(250.8,43.9)(267.4,55.0) \Text(269.1,56.1)[cl]{\BrickRed{$\Psi_R^0$}}
     \SetColor{Black}
     \Vertex(254.5,34.9){1} \Line(254.5,34.9)(274.1,38.8) \Text(276.1,39.2)[cl]{$\chiL$}     
   \end{picture}
\caption{\it Holographic theories for the cases R$_-$ and R$_+$, schematically
represented as a CFT sector coupled to an external source $\Psi_R^0$ (in red) and
to an elementary field $\chiL$ (for $ML>+1/2$).
When both present, $\Psi_R^0$ and $\chiL$ marry through a mass mixing term and 
become heavy. The CFT spectrum is represented by horizontal lines; an isolated line 
indicates a massless chiral bound state. When this state is present, it marries
with $\Psi^0_R$ or $\chiL$.}
\label{fig:Rpm}
}

For a given value $M$ of the bulk mass, we have two possible holographic descriptions,
one in terms of a left-handed source, the other in terms of a right-handed source.
This is similar to the case of the scalar field, where two different holographic theories
exist for certain values of the bulk mass~\cite{Klebanov:1999tb}. 
In our case, however, the two theories
are actually equivalent, being different descriptions of the same physics.
This directly follows by the fact that the equations of motion relate $\Psi_L$ to $\Psi_R$
on the UV boundary, implying that the two possible choices for the source must be
physically equivalent. 

Let us compare the first holographic description, shortly denoted as L$_{\pm}$,
with the second one, R$_\mp$, for a given $M$. For example, we focus on values $ML<-1/2$.
We have $\Sigma_\text{L}\propto f_R/f_L$, and $\Sigma_\text{R}\propto f_L/f_R$, where
$f_{L,R}$ are given by Eq.~(\ref{fRfL}) in the Appendix. 
In absence of additional boundary terms $S_\text{UV}$, 
the equation of motion of the left source reads $\Psi_R^0=0$; we should then
compare a case L$_\pm$ where the left source is dynamical,
with a case R$_\mp$ where the equations of motion force (for example through a 
Lagrange
multiplier) the right-handed source to vanish. This corresponds to an orbifold
theory with parities $\Psi_L(+,\pm)$, $\Psi_R(-,\mp)$.
Hence, the first holographic theory includes a CFT sector coupled to two elementary
fields: the dynamical source $\Psi_L^0$ and the additional field $\chiR$, see Fig.~\ref{fig:Lpm}. 
The second theory is given by a CFT sector probed by an external static source $\Psi_R^0$,
see Fig.~\ref{fig:Rpm}.
The unperturbed spectra of the two CFT sectors are identical: in the limit $\Lambda\to\infty$
the poles of both $\Sigma_L$ and $\Sigma_R$ are given by 
$Y_\beta(m_i L_1)\tan\alpha\pi - J_{\beta}(m_i L_1)=0$. 
For a finite cutoff, the CFT distorted levels satisfy $f_L=0$ ($f_R=0$) in the L$_\pm$
(R$_\mp$) case. In the first theory, the leading correction is of order 
$\Delta m_i \sim m_i\, (m_i/\Lambda)^{2|\alpha|}$; it comes from
the higher-order operator $(\bar{\cal O}_R\parsl {\cal O}_R)/\Lambda^{2|\alpha|}$ and from the 
mixing with the $\chiR$ elementary field. In the second theory, instead, the correction is found 
of order $\Delta m_i\sim m_i\, (m_i/\Lambda)^{2|\alpha-1|}$, as expected from 
the operator $(\bar{\cal O}_L\parsl {\cal O}_L)/\Lambda^{2|\alpha-1|}$.
Even though the two CFT sectors are perturbed by different 
operators and their modified spectrum
is different, the final physical spectrum is the same in both theories.
Indeed, in the first description the source is dynamical and its mixing with the CFT bound states
leads to a physical spectrum given by the zeroes of $\Sigma_L$. 
Thus, in both cases the levels are solutions of $f_L(p)=0$.

\section{Phenomenological applications}
\label{pheap}

The holographic approach described so far is quite powerful in capturing
the essential
qualitative features of the five-dimensional theory.
It is also a useful tool for the computation of physical observables
(at tree level or loop level), since in most of the cases the bulk (or
CFT)
is just a small perturbation to the boundary (or external fields).
For example, for fermions with $|ML|\geq 1/2$, the coupling of the source
to the CFT operator is always, at low-energies, a small parameter.
The same happens with gauge fields.
Therefore, one can perform an expansion
in the boundary-bulk coupling (apart from the ordinary expansion in the 5D
bulk coupling)
that enormously simplify the calculations.
As an illustration, we present here two examples of phenomenological
relevance: Yukawa couplings and  running of the gauge coupling.
Other applications can be found in theories of electroweak symmetry
breaking~\cite{pgb,Burdman:2003ya} or supersymmetry.

Let us consider first the application of holography  to
the calculation of   the Yukawa couplings.
As an example, we consider
 a theory on a slice of 5D AdS, where a scalar $\phi$
and a right-handed quark $q_R$ are confined on the IR boundary, while a 5D
fermion $\Psi=[\Psi_L (++), \Psi_R (--)]$ lives in the bulk.
On the IR boundary one can write the Yukawa interaction term
\begin{equation}
\int\! d^4x \int^{L_1}_{L_0}\!\! dz\, \delta(z-L_1) \sqrt{-g_\text{ind}}
 \left[ \lambda_5\, \bar q_R \phi \Psi_L + h.c. \right]\, ,
\end{equation}
where $\lambda_5$ is a dimensionless coupling.
The size of the physical Yukawa coupling in the 4D low-energy theory
can be easily deduced by looking at the holographic description.
The reader can find a schematic summary of all cases in the last section.
Consider for instance
the description in terms of a left-handed source.
If $ML>-1/2$, the holographic theory is that of Eq.~(\ref{holo}):
an external field $\Psi_L^0$ couples to a CFT sector
with massless bound states $q_R$ and $\phi$.
The Yukawa coupling will be proportional to the physical coupling of
$\Psi_L^0$ to the CFT
evaluated at a low-energy scale $\mu\sim 1/L_1$:
$\lambda(\mu)=(\mu/\Lambda)^{\alpha-1}\, \omega(\mu)/\sqrt{Z_0(\mu)}$.
The physical coupling $\lambda(\mu)$ obeys the RG equation
\begin{equation} \label{RG}
\mu \frac{d \lambda}{d\mu} = \left(\text{Dim}[{\cal O}_R]-\frac{5}{2}\right)
 \lambda + c \frac{N}{16\pi^2}\lambda^3\, ,
\end{equation}
where the second term of the RHS arises from  the 
CFT contribution to the wave-function renormalization
of $\Psi^0_L$ ($c$ is an ${\cal O}(1)$ positive constant). 
Using the relation $\text{Dim}[{\cal O}_R]-5/2=\alpha-1$,
we obtain~\cite{pgb}
\begin{equation}
\lambda(\mu) \sim \frac{4\pi}{\sqrt{N c}}
\left(\frac{\mu}{\Lambda}\right)^{\alpha-1}
 \left[ \frac{\alpha-1}{1- (\mu/\Lambda)^{2\alpha-2}} \right]^{1/2}\, .
\end{equation}
The strength of $\lambda(\mu)$, hence that of the Yukawa coupling, is
determined by the conformal dimension of the operator ${\cal O}_R$.
For $ML>1/2$ ($\alpha>1$), $\Psi_L^0$ has an irrelevant coupling to the CFT,
implying a strong suppression in the physical Yukawa $y$:
$y\sim \lambda(\mu) \sim (\mu/\Lambda)^{\alpha-1} \sqrt{\alpha-1}\,
(4\pi/\sqrt{N})$. If instead $|ML|<1/2$ ($0<\alpha<1$), the coupling is relevant and
$\lambda$ flows to a constant value $\lambda \sim \sqrt{1-\alpha}\, (4\pi/\sqrt{N})$
at low energy.~\footnote{We thank K.~Agashe and R.~Sundrum for having pointed to us the utility
of using the RG equation (\ref{RG}) to deduce the fixed-point value of $\lambda$.}
There is then no suppression in the Yukawa coupling.
This way of realizing the physical Yukawa couplings is quite appealing,
since it allows one to obtain large hierarchies among them in a completely natural
way~\cite{Grossman:1999ra,Gherghetta:2000qt}.
Furthermore, the same mechanism also explains why flavour changing effects
involving
light fermions are suppressed, as far as the only source of flavour
violation are
the couplings on the IR boundary.

Our expectation for the size of the physical Yukawa
can be checked by explicitly computing the low-energy effective
theory following the holographic approach:
one can integrate out the bulk dynamics and write an effective boundary
lagrangian on the UV boundary. The 4D low-energy degrees of freedom are the
elementary fields plus eventual massless CFT composites,
like $q_R$ and $\phi$ in our case, that will be treated separately from
the other massive CFT bound states.
The Green functions of the boundary theory can be conveniently computed in
terms of 5D bulk correlators with external legs attached on the UV boundary.
For example, the 5D fermion propagator
with end points on the UV boundary corresponds to the inverse of the
quadratic term in the boundary lagrangian: $S(p,L_0,L_0)=\Sigma^{-1}(p)$.
The physical Yukawa is given in terms of a 5D propagator
from the UV boundary, where $\Psi_L^0$ lives, to the IR boundary, where $q_R$,
$\phi$ are confined:
\begin{equation} \label{eq:holoYuk}
y = \lim_{p\to 0}\, \lambda_5\, \frac{S(p,L_0,L_1)}{S(p,L_0,L_0)}\, \,
 \frac{(L/L_1)^{3/2}}{\sqrt{Z_0(\mu)}}
  = g_5\, \lambda_5 \left(\frac{L_0}{L_1}\right)^{\alpha-1}\!
   \left(\frac{2\alpha-2}{1-(L_0/L_1)^{2\alpha-2}}\right)^{1/2}, \quad
\alpha>0\, .
\end{equation}
We have divided by $S(p,L_0,L_0)$ to obtain the amputated Green function,
and taken into account a factor $(L/L_1)^{3/2}/\sqrt{Z_0(\mu)}$
that appears after all fields are canonically normalized.
Here $Z_0(\mu)=L/g_5^2 \left(1-(L_0/L_1)^{2\alpha-2}\right)/(2\alpha-2)$,
is
the low-energy wave function of the elementary field $\Psi_L^0$, as
extracted
from $S(p,L_0,L_0)$ (or equivalently from $\Sigma^{-1}(p)$)
in the limit of small 4D momentum, $p\ll 1/L_1$.
As expected, there is a suppression in the physical Yukawa (only) for
$\alpha>1$.
In the case of bulk mass $ML<-1/2$ ($\alpha<0$), the holographic
description changes (see the figures of the last section)
and the left-handed massless quark is a bound state of the CFT.
Being a coupling among three composites, the Yukawa is expected to have
no suppression whatsoever.
This result coincides with the calculation in the KK  approach.

As a second example, we consider the running of gauge couplings in a Grand
Unified theory (GUT) defined on a slice of 5D AdS.
The 5D one-loop correction to gauge couplings has a holographic
interpretation
as a CFT contribution plus a contribution from a loop of elementary
(external) fields (see for instance~\cite{Contino:2002kc,Goldberger:2003mi}).
If the GUT symmetry is broken on the UV boundary but not in the bulk,
the only differential running in the gauge couplings comes from the
elementary modes.
Consider for instance an SU(5) group in the bulk,
reduced by boundary conditions to SU(3)$\times$SU(2)$\times$U(1) on the
UV boundary.
Whether the SU(5) symmetry is broken or not on the IR boundary
is not relevant for the evolution of the gauge couplings at energies
$E>1/L_1$, as obvious from the holography picture.
Let us also introduce a fermion $\Psi$ in the bulk,
transforming as a \textbf{5} of SU(5).
We can assign the following UV boundary conditions:
\begin{equation}
\Psi = \begin{bmatrix} \Psi^{(D)}_L (+) &  \Psi^{(D)}_R (-) \\[0.15cm]
                       \Psi^{(T)}_L (-) &  \Psi^{(T)}_R (+)
\end{bmatrix}\, ,
\end{equation}
where the components $\Psi^{(D)}$ and $\Psi^{(T)}$ transform respectively as
a doublet of the SU(2) and a triplet of the SU(3) subgroup.
The holographic theory consists
in a CFT with a global SU(5) symmetry whose
SU(3)$\times$SU(2)$\times$U(1) subgroup is gauged by external vector
fields.
The fermionic content of the holographic theory depends on the bulk mass
$M$ of the
5D field $\Psi$ (see last section for a summary).
If $ML>1/2$ ($\alpha >1$), the theory only has  a light 
elementary fermion $\psi_L^{(D)}$.
It has an irrelevant coupling to the CFT and it contributes to the
differential running
with the usual one-loop beta function of an SU(2) doublet.
If instead $|ML|<1/2$ ($0<\alpha <1$), the coupling between $\psi_L^{(D)}$
and the CFT becomes relevant. In this case we expect that
the contribution of $\psi_L^{(D)}$ to the running is largely deformed
due to the strong interaction with the CFT.
Finally, if $ML<-1/2$ ($\alpha <0$),
the holographic theory changes: there is a light elementary
fermion $\psi_R^{(T)}$, transforming now as a triplet of SU(3),
with an irrelevant coupling to the CFT.
We then expect a large logarithm in the differential running,
this time with the beta function of a triplet of SU(3).
This holographic prediction agrees with the explicit 5D one-loop
computation
performed in Ref.~\cite{Choi:2002ps} using the KK approach.

Summarizing: for bulk masses $|ML|>1/2$ the holographic description of the
5D theory
predicts large logs in the difference of gauge couplings at low energy.
This has a phenomenological interest since it shows that a differential
effect can be
obtained from fermions, even though 5D fields come in complete SU(5)
multiplets.
Holography is  very useful  since it  exactly predicts the coefficient of
the
logarithm,
relating it to the beta-function of the elementary modes.
Notice that it is really the elementary field content, rather than the
zero-mode KK spectrum, which determines the effect.

\section{Conclusions: the dictionary}

We have presented the holographic description
of a theory with  fermion living in a slice of 5D AdS.
In particular, we have shown  how two equivalent descriptions can be formulated,
using either $\Psi^0_L$ or $\Psi^0_R$ as the external source.
Here we summarize these results by giving 
the dictionary which relates the orbifold 5D AdS theory with the 4D CFT.
We hope that this can be a useful resource for model builders,
who can quickly deduce the 4D holographic description of their 5D theory once
the boundary conditions and the bulk mass of the 5D fermion field are specified.
All the cases are summarized by the pictures in the following two pages, where
the same notation of the text has been followed. 
We present four different parity assignments for the 5D fermion $\Psi_L$:
  ($+,+$), ($+,-$),  ($-,+$) and ($-,-$).   
For each case
we present the two equivalent 
4D holographic descriptions with their corresponding  4D lagrangian.
No extra UV boundary terms have been considered, and  nondynamical
external fields are not depicted.

\newpage
\vspace*{-1cm}
{\small
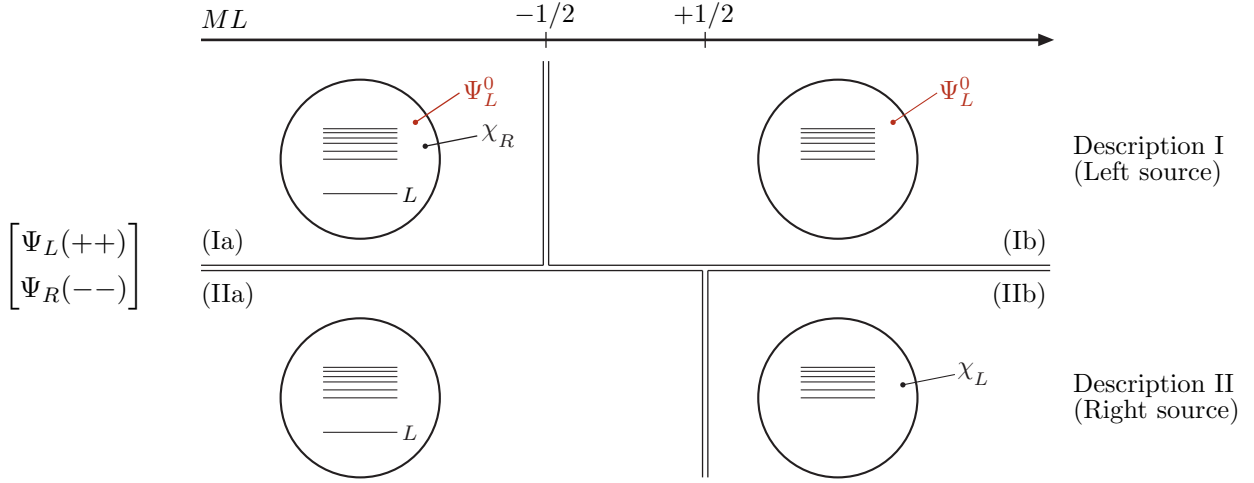
\centering
   \begin{picture}(300,200)
     \Text(-85,90)[lc]{\normalsize $\begin{bmatrix} \Psi_L(++) \\[0.1cm] \Psi_R (--) \end{bmatrix}$}
     \Text(320,135)[lc]{Description I} \Text(320,125)[lc]{(Left source)}
     \Text(320,45)[lc]{Description II} \Text(320,35)[lc]{(Right source)}
     \SetWidth{1} \LongArrow(-10,175)(310,175) \SetWidth{0.5}
     \Line(120,172)(120,178) \Line(180,172)(180,178) 
     \Text(120,180)[cb]{$-1/2$} \Text(180,180)[cb]{$+1/2$} \Text(-10,180)[lb]{$ML$}
     \Line(119,167)(119,90) \Line(121,167)(121,90) 
     \Line(121,90)(310,90) \Line(119,90)(-10,90) \Line(-10,88)(179,88) \Line(181,88)(310,88)
     \Line(179,88)(179,10) \Line(181,88)(181,10)
     \Text(-10,94)[bl]{(Ia)}   \Text(-10,84)[tl]{(IIa)} 
     \Text(310,94)[br]{(Ib)}  \Text(310,84)[tr]{(IIb)}
     \SetWidth{0.8} \CArc(50,130)(30,0,360) \SetWidth{0.4} 
     \Line(36,130)(64,130) \Line(36,133)(64,133) \Line(36,136)(64,136) 
     \Line(36,138)(64,138) \Line(36,140)(64,140) \Line(36,141.5)(64,141.5) 
     \Line(36,117)(64,117) \Text(66,117)[lc]{\scriptsize $L$} \SetWidth{0.5} 
     \SetColor{BrickRed} 
     \Vertex(70.8,143.9){1} \Line(70.8,143.9)(87.4,155.0) \Text(89.1,156.1)[cl]{\BrickRed{$\Psi_L^0$}}
     \SetColor{Black}
     \Vertex(74.5,134.9){1} \Line(74.5,134.9)(94.1,138.8) \Text(96.1,139.2)[cl]{$\chiR$}
     \SetWidth{0.8} \CArc(230,130)(30,0,360) \SetWidth{0.4} 
     \Line(216,130)(244,130) \Line(216,133)(244,133) \Line(216,136)(244,136) 
     \Line(216,138)(244,138) \Line(216,140)(244,140) \Line(216,141.5)(244,141.5) 
     \SetColor{BrickRed} 
     \Vertex(250.8,143.9){1} \Line(250.8,143.9)(267.4,155.0) 
     \Text(269.1,156.1)[cl]{\BrickRed{$\Psi_L^0$}}
     \SetColor{Black}
     \SetWidth{0.8} \CArc(50,40)(30,0,360) \SetWidth{0.4} 
     \Line(36,40)(64,40) \Line(36,43)(64,43) \Line(36,46)(64,46) 
     \Line(36,48)(64,48) \Line(36,50)(64,50) \Line(36,51.5)(64,51.5) 
     \Line(36,27)(64,27) \Text(66,27)[lc]{\scriptsize $L$} \SetWidth{0.5} 
     \SetWidth{0.8} \CArc(230,40)(30,0,360) \SetWidth{0.4} 
     \Line(216,40)(244,40) \Line(216,43)(244,43) \Line(216,46)(244,46) 
     \Line(216,48)(244,48) \Line(216,50)(244,50) \Line(216,51.5)(244,51.5) 
     \SetColor{Black}
     \Vertex(254.5,44.9){1} \Line(254.5,44.9)(274.1,48.8) \Text(276.1,49.2)[cl]{$\chiL$}   
   \end{picture} 

{\normalsize
\begin{equation*}
\begin{split}
{\cal L}_\text{Ia}=& {\cal L}_{\rm CFT} + Z_0\, \bar\Psi^0_L i\!\parsl\Psi_L^0
 + \tilde Z_0\, \chiRbar i\!\parsl\chiR 
 +  \eta\Lambda \left( \chiRbar \Psi^0_L +\bar\Psi^0_L \chiR \right) \\
 &+ \left[\frac{\omega}{\Lambda^{|\alpha|-1}}\, \bar\Psi^0_L{\cal O}_R
 + \frac{\tilde\omega}{\Lambda^{|\alpha|}}\, \chiRbar\parsl{\cal O}_R+ h.c.\right]
 + \xi\,\frac{\bar{\cal  O}_R\parsl{\cal O}_R}{\Lambda^{2|\alpha|}} \\[0.3cm]
{\cal L}_\text{Ib}=&{\cal L}_{\rm CFT} + Z_0\, \bar\Psi^0_L i\!\parsl\Psi^0_L 
 + \frac{\omega}{\Lambda^{\alpha-1}} \left(\bar\Psi^0_L{\cal O}_R+h.c.\right)
 + \xi\, \frac{\bar{\cal  O}_R\parsl{\cal O}_R}{\Lambda^{2\alpha}} \\[0.3cm]
{\cal L}_\text{IIa}=& {\cal L}_{\rm CFT} 
 + \xi\, \frac{\bar{\cal  O}_L\parsl{\cal O}_L}{\Lambda^{2|\alpha-1|}} \\[0.3cm]
{\cal L}_\text{IIb}=&{\cal L}_{\rm CFT} + \tilde Z_0\, \chiLbar i\!\parsl\chiL
 + \frac{\tilde\omega}{\Lambda^{\alpha-1}} \left(\chiLbar\parsl {\cal O}_L+h.c.\right) 
 + \xi\, \frac{\bar{\cal O}_L\parsl {\cal O}_L}{\Lambda^{2(\alpha-1)}} 
\end{split}
\end{equation*}
}

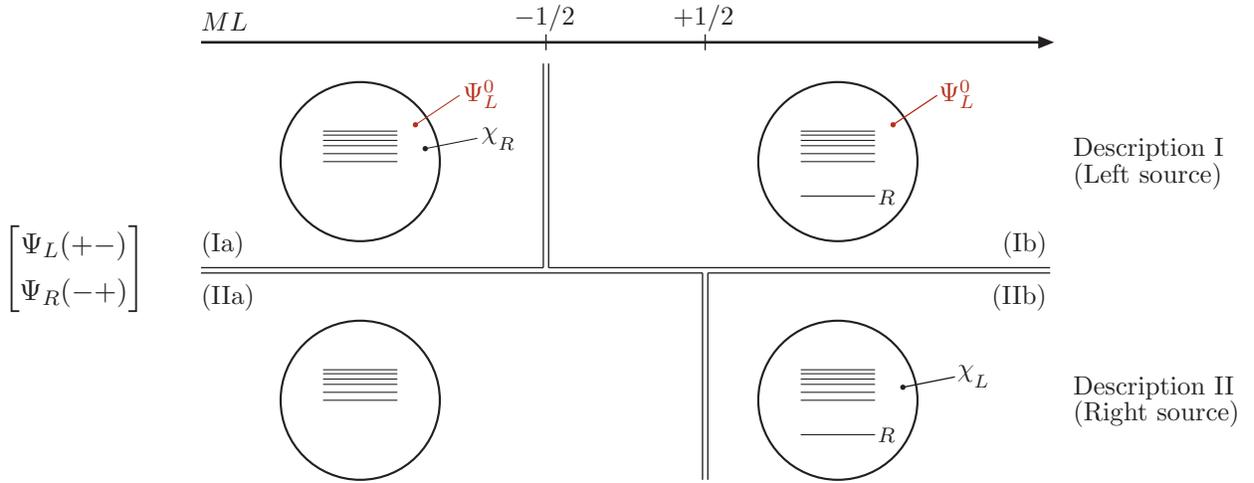
\vspace*{0.5cm}
   \begin{picture}(300,200)
     \Text(-85,90)[lc]{\normalsize $\begin{bmatrix} \Psi_L(+-) \\[0.1cm] \Psi_R (-+) \end{bmatrix}$}
     \Text(320,135)[lc]{Description I} \Text(320,125)[lc]{(Left source)}
     \Text(320,45)[lc]{Description II} \Text(320,35)[lc]{(Right source)}
     \SetWidth{1} \LongArrow(-10,175)(310,175) \SetWidth{0.5}
     \Line(120,172)(120,178) \Line(180,172)(180,178) 
     \Text(120,180)[cb]{$-1/2$} \Text(180,180)[cb]{$+1/2$} \Text(-10,180)[lb]{$ML$}
     \Line(119,167)(119,90) \Line(121,167)(121,90) 
     \Line(121,90)(310,90) \Line(119,90)(-10,90) \Line(-10,88)(179,88) \Line(181,88)(310,88)
     \Line(179,88)(179,10) \Line(181,88)(181,10)
     \Text(-10,94)[bl]{(Ia)}  \Text(-10,84)[tl]{(IIa)} 
     \Text(310,94)[br]{(Ib)}  \Text(310,84)[tr]{(IIb)}
     \SetWidth{0.8} \CArc(50,130)(30,0,360) \SetWidth{0.4} 
     \Line(36,130)(64,130) \Line(36,133)(64,133) \Line(36,136)(64,136) 
     \Line(36,138)(64,138) \Line(36,140)(64,140) \Line(36,141.5)(64,141.5) 
     \SetColor{BrickRed} 
     \Vertex(70.8,143.9){1} \Line(70.8,143.9)(87.4,155.0) \Text(89.1,156.1)[cl]{\BrickRed{$\Psi_L^0$}}
     \SetColor{Black}
     \Vertex(74.5,134.9){1} \Line(74.5,134.9)(94.1,138.8) \Text(96.1,139.2)[cl]{$\chiR$}
     \SetWidth{0.8} \CArc(230,130)(30,0,360) \SetWidth{0.4} 
     \Line(216,130)(244,130) \Line(216,133)(244,133) \Line(216,136)(244,136) 
     \Line(216,138)(244,138) \Line(216,140)(244,140) \Line(216,141.5)(244,141.5) 
     \Line(216,117)(244,117) \Text(246,117)[lc]{\scriptsize $R$} \SetWidth{0.5} 
     \SetColor{BrickRed} 
     \Vertex(250.8,143.9){1} \Line(250.8,143.9)(267.4,155.0) 
     \Text(269.1,156.1)[cl]{\BrickRed{$\Psi_L^0$}}
     \SetColor{Black}
     \SetWidth{0.8} \CArc(50,40)(30,0,360) \SetWidth{0.4} 
     \Line(36,40)(64,40) \Line(36,43)(64,43) \Line(36,46)(64,46) 
     \Line(36,48)(64,48) \Line(36,50)(64,50) \Line(36,51.5)(64,51.5) 
     \SetWidth{0.8} \CArc(230,40)(30,0,360) \SetWidth{0.4} 
     \Line(216,40)(244,40) \Line(216,43)(244,43) \Line(216,46)(244,46) 
     \Line(216,48)(244,48) \Line(216,50)(244,50) \Line(216,51.5)(244,51.5) 
     \Line(216,27)(244,27) \Text(246,27)[lc]{\scriptsize $R$} \SetWidth{0.5} 
     \SetColor{Black}
     \Vertex(254.5,44.9){1} \Line(254.5,44.9)(274.1,48.8) \Text(276.1,49.2)[cl]{$\chiL$}   
   \end{picture} \\[0.5cm]

\newpage
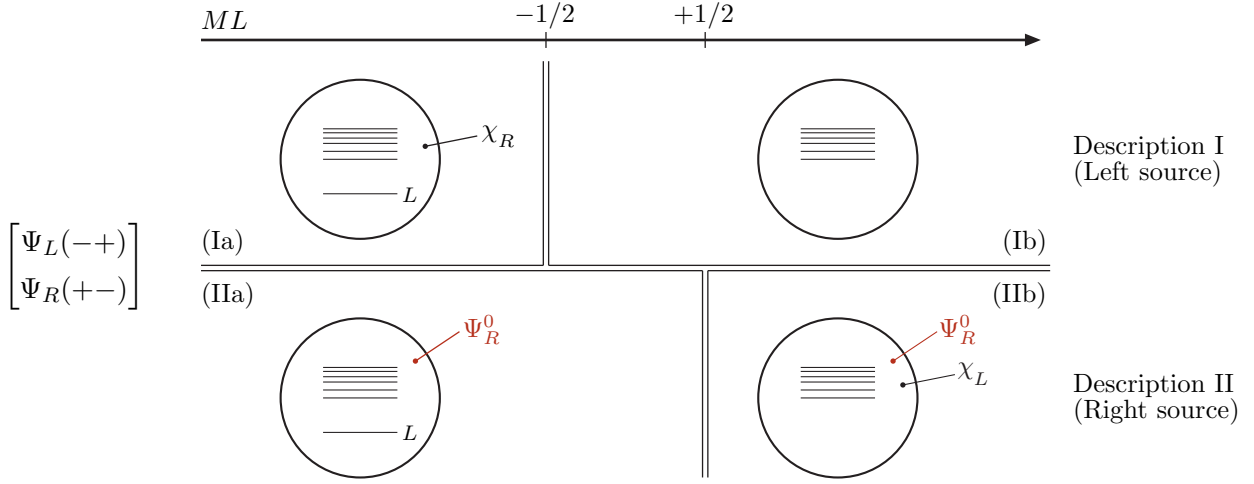
\vspace*{-1cm}
   \begin{picture}(300,200)
     \Text(-85,90)[lc]{\normalsize $\begin{bmatrix} \Psi_L(-+) \\[0.1cm] \Psi_R (+-) \end{bmatrix}$}
     \Text(320,135)[lc]{Description I} \Text(320,125)[lc]{(Left source)}
     \Text(320,45)[lc]{Description II} \Text(320,35)[lc]{(Right source)}
     \SetWidth{1} \LongArrow(-10,175)(305,175) \SetWidth{0.5}
     \Line(120,172)(120,178) \Line(180,172)(180,178) 
     \Text(120,180)[cb]{$-1/2$} \Text(180,180)[cb]{$+1/2$} \Text(-10,180)[lb]{$ML$}
     \Line(119,167)(119,90) \Line(121,167)(121,90) 
     \Line(121,90)(310,90) \Line(119,90)(-10,90) \Line(-10,88)(179,88) \Line(181,88)(310,88)
     \Line(179,88)(179,10) \Line(181,88)(181,10)
     \Text(-10,94)[bl]{(Ia)}   \Text(-10,84)[tl]{(IIa)} 
     \Text(310,94)[br]{(Ib)}  \Text(310,84)[tr]{(IIb)}
     \SetWidth{0.8} \CArc(50,130)(30,0,360) \SetWidth{0.4} 
     \Line(36,130)(64,130) \Line(36,133)(64,133) \Line(36,136)(64,136) 
     \Line(36,138)(64,138) \Line(36,140)(64,140) \Line(36,141.5)(64,141.5) 
     \Line(36,117)(64,117) \Text(66,117)[lc]{\scriptsize $L$} \SetWidth{0.5} 
     \SetColor{Black}
     \Vertex(74.5,134.9){1} \Line(74.5,134.9)(94.1,138.8) \Text(96.1,139.2)[cl]{$\chiR$}
     \SetWidth{0.8} \CArc(230,130)(30,0,360) \SetWidth{0.4} 
     \Line(216,130)(244,130) \Line(216,133)(244,133) \Line(216,136)(244,136) 
     \Line(216,138)(244,138) \Line(216,140)(244,140) \Line(216,141.5)(244,141.5) 
     \SetWidth{0.8} \CArc(50,40)(30,0,360) \SetWidth{0.4} 
     \Line(36,40)(64,40) \Line(36,43)(64,43) \Line(36,46)(64,46) 
     \Line(36,48)(64,48) \Line(36,50)(64,50) \Line(36,51.5)(64,51.5) 
     \Line(36,27)(64,27) \Text(66,27)[lc]{\scriptsize $L$} \SetWidth{0.5} 
     \SetColor{BrickRed} 
     \Vertex(70.8,53.9){1} \Line(70.8,53.9)(87.4,65.0) \Text(89.1,66.1)[cl]{\BrickRed{$\Psi_R^0$}}
     \SetColor{Black}
     \SetWidth{0.8} \CArc(230,40)(30,0,360) \SetWidth{0.4} 
     \Line(216,40)(244,40) \Line(216,43)(244,43) \Line(216,46)(244,46) 
     \Line(216,48)(244,48) \Line(216,50)(244,50) \Line(216,51.5)(244,51.5) 
     \SetColor{BrickRed} 
     \Vertex(250.8,53.9){1} \Line(250.8,53.9)(267.4,65.0) \Text(269.1,66.1)[cl]{\BrickRed{$\Psi_R^0$}}
     \SetColor{Black}
     \Vertex(254.5,44.9){1} \Line(254.5,44.9)(274.1,48.8) \Text(276.1,49.2)[cl]{$\chiL$}   
   \end{picture} 

{\normalsize
\begin{equation*}
\begin{split}
 {\cal L}_\text{Ia}=&{\cal L}_{\rm CFT} + \tilde Z_0\, \chiRbar i\!\parsl\chiR 
  + \frac{\tilde\omega}{\Lambda^{|\alpha|}} \left(\chiRbar\parsl {\cal O}_R+h.c.\right) 
  + \xi\, \frac{\bar{\cal O}_R\parsl {\cal O}_R}{\Lambda^{2|\alpha|}} \\[0.3cm]
 {\cal L}_\text{Ib}=& {\cal L}_{\rm CFT} 
  + \xi\, \frac{\bar{\cal  O}_R \parsl{\cal O}_R}{\Lambda^{2 \alpha}} \\[0.3cm]
 {\cal L}_\text{IIa}=&{\cal L}_{\rm CFT} + Z_0\, \bar\Psi^0_R i\!\parsl\Psi^0_R
  + \frac{\omega}{\Lambda^{-\alpha}} \left(\bar\Psi^0_R {\cal O}_L+h.c.\right)
  + \xi\, \frac{\bar{\cal  O}_L\parsl{\cal O}_L}{\Lambda^{2|\alpha-1|}} \\[0.3cm]
 {\cal L}_\text{IIb}=& {\cal L}_{\rm CFT} + Z_0\, \bar\Psi^0_R i\!\parsl\Psi_R^0
  + \tilde Z_0\, \chiLbar i\!\parsl\chiL 
  +  \eta\Lambda \left( \chiLbar \Psi^0_R +\bar\Psi^0_R \chiL \right) \\ 
  &+ \left[\frac{\omega}{\Lambda^{\alpha-2}}\, \bar\Psi^0_R{\cal O}_L
  + \frac{\tilde\omega}{\Lambda^{\alpha-1}}\, \chiLbar\parsl{\cal O}_L+ h.c.\right]
  + \xi\,\frac{\bar{\cal  O}_L\parsl{\cal O}_L}{\Lambda^{2(\alpha-1)}}
\end{split}
\end{equation*}
}

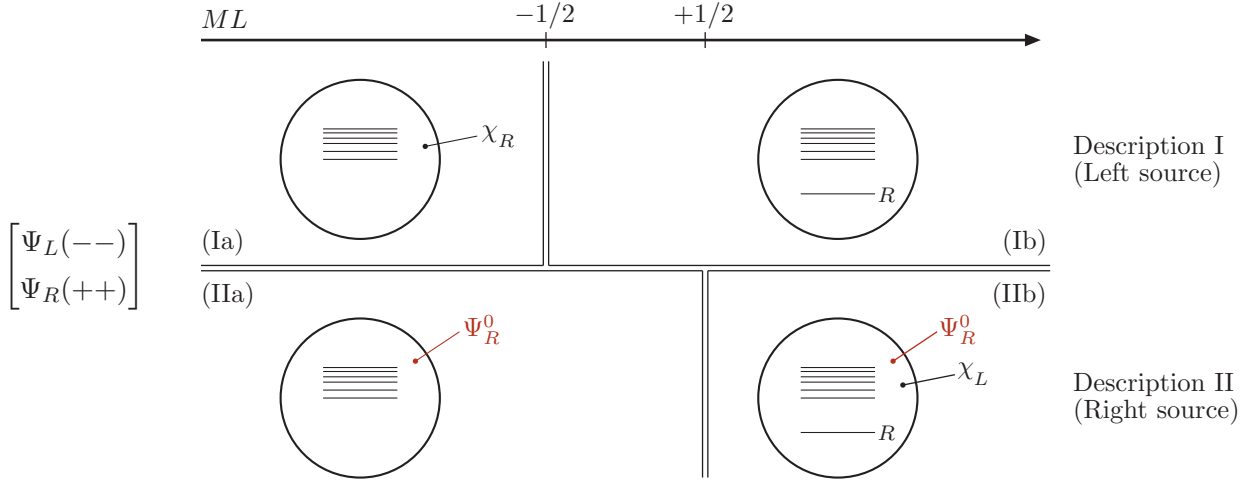
\vspace*{0.5cm}
   \begin{picture}(300,200)
     \Text(-85,90)[lc]{\normalsize $\begin{bmatrix} \Psi_L(--) \\[0.1cm] \Psi_R (++) \end{bmatrix}$}
     \Text(320,135)[lc]{Description I} \Text(320,125)[lc]{(Left source)}
     \Text(320,45)[lc]{Description II} \Text(320,35)[lc]{(Right source)}
     \SetWidth{1} \LongArrow(-10,175)(305,175) \SetWidth{0.5}
     \Line(120,172)(120,178) \Line(180,172)(180,178) 
     \Text(120,180)[cb]{$-1/2$} \Text(180,180)[cb]{$+1/2$} \Text(-10,180)[lb]{$ML$}
     \Line(119,167)(119,90) \Line(121,167)(121,90) 
     \Line(121,90)(310,90) \Line(119,90)(-10,90) \Line(-10,88)(179,88) \Line(181,88)(310,88)
     \Line(179,88)(179,10) \Line(181,88)(181,10)
     \Text(-10,94)[bl]{(Ia)}  \Text(-10,84)[tl]{(IIa)} 
     \Text(310,94)[br]{(Ib)}  \Text(310,84)[tr]{(IIb)}
     \SetWidth{0.8} \CArc(50,130)(30,0,360) \SetWidth{0.4} 
     \Line(36,130)(64,130) \Line(36,133)(64,133) \Line(36,136)(64,136) 
     \Line(36,138)(64,138) \Line(36,140)(64,140) \Line(36,141.5)(64,141.5) 
     \SetColor{Black}
     \Vertex(74.5,134.9){1} \Line(74.5,134.9)(94.1,138.8) \Text(96.1,139.2)[cl]{$\chiR$}
     \SetWidth{0.8} \CArc(230,130)(30,0,360) \SetWidth{0.4} 
     \Line(216,130)(244,130) \Line(216,133)(244,133) \Line(216,136)(244,136) 
     \Line(216,138)(244,138) \Line(216,140)(244,140) \Line(216,141.5)(244,141.5) 
     \Line(216,117)(244,117) \Text(246,117)[lc]{\scriptsize $R$} \SetWidth{0.5} 
     \SetColor{Black}
     \SetWidth{0.8} \CArc(50,40)(30,0,360) \SetWidth{0.4} 
     \Line(36,40)(64,40) \Line(36,43)(64,43) \Line(36,46)(64,46) 
     \Line(36,48)(64,48) \Line(36,50)(64,50) \Line(36,51.5)(64,51.5) 
     \SetColor{BrickRed} 
     \Vertex(70.8,53.9){1} \Line(70.8,53.9)(87.4,65.0) \Text(89.1,66.1)[cl]{\BrickRed{$\Psi_R^0$}}
     \SetColor{Black}
     \SetWidth{0.8} \CArc(230,40)(30,0,360) \SetWidth{0.4} 
     \Line(216,40)(244,40) \Line(216,43)(244,43) \Line(216,46)(244,46) 
     \Line(216,48)(244,48) \Line(216,50)(244,50) \Line(216,51.5)(244,51.5) 
     \Line(216,27)(244,27) \Text(246,27)[lc]{\scriptsize $R$} \SetWidth{0.5} 
     \SetColor{BrickRed} 
     \Vertex(250.8,53.9){1} \Line(250.8,53.9)(267.4,65.0) \Text(269.1,66.1)[cl]{\BrickRed{$\Psi_R^0$}}
     \SetColor{Black}
     \Vertex(254.5,44.9){1} \Line(254.5,44.9)(274.1,48.8) \Text(276.1,49.2)[cl]{$\chiL$}   
   \end{picture} \\[0.5cm]
}

\acknowledgments
R.~C. thanks Kaustubh Agashe, Michele Redi and Raman Sundrum for useful discussions.
The work of A.~P. was  supported  in part by the MCyT and FEDER Research Project
FPA2002-00748 and DURSI Research Project 2001-SGR-00188.
The work of R.~C. is supported by NSF grants P420D3620414350 and P420D3620434350.

\appendix
\section{Functions $f_{L,R}$ for the AdS case}

In this appendix we give the functions $f_{L,R}$ for the case of
a fermion field living on a slice of 5D AdS space. 
They are defined to be solutions of the system of equations (\ref{dirac5d}), 
where $a(z)=L/z$ in the specific case of AdS. 
This system of first-order coupled equations can be translated into two separate 
second-order equations for $f_L$ and $f_R$:
\begin{equation} \label{eq:2ndorder}
\left[ \partial_z^2 -\frac{4}{z}\partial_z + p^2 + \frac{6}{z^2} \mp \frac{ML}{z^2}
 - \frac{(ML)^2}{z^2} \right] f_{L,R}(p,z) = 0\, .
\end{equation}
It is convenient to compute $f_L$ or $f_R$ from (\ref{eq:2ndorder}) and then solve for the other 
function using the system of linear equations. In this way $f_L$, $f_R$ are
correctly normalized up to a common factor $N(p,L_1)$ which, consistently with 
Eq.~(\ref{sollr}), 
can be chosen to be $N(p,L_1)=1$. The final result 
is given by~\cite{Grossman:1999ra,Gherghetta:2000qt}~\footnote{
It is commonly used in the literature a different basis of Bessel functions,
$\{ J_{\alpha_{L,R}},Y_{\alpha_{L,R}} \}$, whose index is defined to be strictly positive,
$\alpha_{L,R}\equiv |ML\pm 1/2|$; see for instance \cite{Gherghetta:2000qt}.
However, expressions are more compact if one allows the indeces to be 
negative, as assumed in Eq.~(\ref{fRfL}).
We have explicitly checked that the two choices are completely equivalent.}
\begin{equation} \label{fRfL}
 \begin{split}
 f_L(p,z) &= z^{5/2} 
  \left[ J_\alpha(pz)Y_{\beta}(pL_1) - J_{\beta}(pL_1)Y_{\alpha}(pz) \right] \\
 f_R(p,z) &= z^{5/2} 
  \left[ J_{\alpha-1}(pz)Y_{\beta}(pL_1) - J_{\beta}(pL_1)Y_{\alpha-1}(pz) \right]\, ,
 \end{split}   
\end{equation}
where $\alpha\equiv ML+1/2$ and $\beta=\alpha-1$ for $\Psi_R(L_1)=0$, 
$\beta=\alpha$ for $\Psi_L(L_1)=0$.
The effective boundary action (\ref{sigma}) immediately follows from Eq.~(\ref{fRfL}).

\newpage


\end{document}